\documentclass{aa}
\usepackage{epsfig}
\usepackage{natbib}
\usepackage{graphicx}
\usepackage{txfonts}
\begin{document}
   \title{Absence of significant cross-correlation between 
   WMAP and SDSS}

   \subtitle{}

   \author{M. L\'opez-Corredoira\inst{1,2}, F. Sylos
   Labini\inst{3,4}, J. Betancort-Rijo\inst{1,2}}

   \offprints{martinlc@iac.es}

\institute{
$^1$ Instituto de Astrof\'\i sica de Canarias,
E-38200 La Laguna, Tenerife, Spain\\
$^2$ Departamento de Astrof\'\i sica, Universidad de La Laguna,
E-38205 La Laguna, Tenerife, Spain\\
$^3$ Centro Studi e Ricerche Enrico Fermi, Via Panisperna 89 A, Compendio del
Viminale, 00184 Rome, Italy\\
$^4$ Istituto dei Sistemi Complessi CNR, Via dei Taurini 19, 00185 Rome,
Italy}

   \date{Received xxxx; accepted xxxx}

 
  \abstract
  {}      
%
{Several authors have claimed to detect a significant
  cross-correlation between microwave WMAP anisotropies and the SDSS
  galaxy distribution.  We repeat these analyses to determine the different
  cross-correlation uncertainties caused by re-sampling errors and field-to-field
  fluctuations. The first type of error concerns
  overlapping sky regions, while the second type concerns non-overlapping
  sky regions.} 
%
  {To measure the re-sampling errors, we use bootstrap and jack-knife
      techniques. For the field-to-field fluctuations, we use 
      three methods: 1) evaluation of the dispersion in the
    cross-correlation when correlating separated
    regions of WMAP with the original region of SDSS; 
    2) use of mock Monte Carlo WMAP maps;
    3) a new method (developed in this article), which measures the error
    as a function of the integral of the product of the self-correlations for each map.}
%
  {The average cross-correlation for $b>30$ deg. is
    significantly stronger than the re-sampling errors---both the jack-knife
    and bootstrap techniques provide similar results---but it is of the order of
    the field-to-field fluctuations. This is confirmed by the
    cross-correlation between anisotropies and galaxies in more
    than the half of the sample being null within re-sampling errors.} 
%
{Re-sampling methods underestimate the errors. 
Field-to-field fluctuations dominate the 
  detected signals. The ratio of signal to re-sampling 
  errors is larger than unity in a way that strongly depends on the 
  selected sky region. We therefore conclude that there is no 
  evidence yet of a significant detection of the integrated Sachs-Wolfe 
  (ISW) effect. 
  Hence, the value of $\Omega _\Lambda\approx 0.8$ obtained 
  by the authors who assumed they were observing the ISW effect 
  would appear to have originated from noise analysis.}

\titlerunning{Cross-correlation of WMAP vs SDSS}
\authorrunning{L\'opez-Corredoira et al.}

\maketitle
%

\section{Introduction}

Several authors
\citep{Fosalba03,Vielva06,Cabre06,Raccanelli08,Ho08,Granett08} have 
claimed that there is a significant cross-correlation between cosmic 
microwave background radiation (CMBR) anisotropies and the density of 
galaxies, which
is interpreted as the integrated Sachs-Wolfe (ISW) effect. An anticorrelation caused
by the Sunyaev-Zel'dovich effect would also be expected on scales smaller
than $\sim 1^\circ $, but this is negligible when averaging large
regions of the sky \citep{Hernandez04}.
The conclusion of these authors is that the measured 
cross-correlation should be interpreted as a detection of the ISW effect
within a $\Lambda $CDM-cosmology 
and it serves to constrain the value of the cosmological parameters.

We reanalyze whether this correlation exists by considering 
galaxies observed by the Sloan Digital Sky Survey (SDSS), 
taking particular care in the
calculation of the cross-correlation errors. 
The root mean square (r.m.s.) of the cross-correlation for distant, widely different
areas of sky (here called ``field-to-field'' errors) 
infer much larger errors than those calculated
 using re-sampling cross-correlations techniques i.e., when 
 these are determined in different strongly overlapping and thus not 
 independent subsamples of a given sample (re-sampling errors).
We conclude that measurements of the errors in the cross-correlation 
function for overlapping
sub-fields lead to an underestimate of the true scatter in the signal.

\section{Data}

We consider two types of data for the
two fields that we cross-correlate:

\begin{enumerate}

\item Microwave temperature anisotropies ($\delta T$) from the 5th year WMAP
  release \citep{Hinshaw09}. We use the V-band (61 GHz)
  data because of its lower level of pixel noise. We checked that
  the results of this paper are approximately similar if we use the
  W-band (94 GHz) data.  There is no need to subtract foreground
  Galactic contamination because this is not correlated with galaxy
  counts (corrected for extinction), and because this is small in
  off-plane regions. In any case, the published foreground corrections
  might not be enough accurate \citep{LopezCorredoira07}.
  We assign the same weight to each WMAP pixel of equal size.

\item Galaxy counts ($G$) are obtained from the survey SDSS,
  photometric catalog, data release DR7 \citep{Abazajian09}. They
  cover an area 11,663 deg$^2$ (28\% of the sky) mostly in the northern
  Galactic hemisphere.  We did not use the striped region data with $b<30^\circ $ to
ensure low Galactic extinction and avoided
  negative latitudes because these are small isolated regions dominated
  by edge effects. We used only galaxies with $r$ magnitudes in the
  range $[18,21]$ (Galactic-extinction corrected) (within these limits
  galaxy counts are complete) and ``clean photometry'' according to
  an SDSS algorithm (e.g., we removed sources close to saturated objects
  with contamination of their by other objects), and avoiding the borders
  by $\sim$ 0.3 deg. In this situation, the total used area is
  7,349 deg$^2$ (18\% of the sky), containing 2.2, 6.4 and 17.1 million
  galaxies in the $r$ magnitude ranges [18,19], [19,20], and [20,21],
  respectively.

\end{enumerate}

\section{Methods}
\label{.methods}

By defining the galaxy count ($G$) density contrast to be
$\delta_G(\theta) = (G(\theta) - \langle G \rangle) / \langle G
\rangle$, and denoting by $\delta_T(\theta) = T(\theta) - T_0$ the
fluctuations in the CMBR with respect to the average temperature
$T_0$, the cross-correlation function can be written as
\begin{equation}
\omega _{TG}(\theta) \equiv \langle \delta_T(\theta) \delta_G(0) \rangle 
.\label{crosscorr}
\end{equation} 
The estimator of Eq. (\ref{crosscorr}) computes the cross-correlation
to be the average over all pixels with separations $\theta \pm (\Delta
\theta /2)$, where $\Delta \theta $ is the step between successive
values of $\theta $.  In what follows, we set $\Delta \theta
  \approx 0.29$ deg.

There are two kinds of errors 
in the cross-correlation, associated with two distinct ways of
constructing sub-fields over which they are computed
\citep{SylosLabini09}:
\begin{enumerate}

\item Re-sampling errors: for point distributions, there
is a component of the total error that is caused by the finiteness of the 
number of points and is closely related to that given by the
re-sampling techniques \cite{Betancort-Rijo91};
however, here, in the correlation of two continuous fields, the association
is not at all clear. These may be estimated with a
  re-sampling technique, for instance jack-knife or bootstrap. In the
  latter case, we calculate $n_s$ times the cross-correlation by
  removing each time a different fraction $1/n_s$ of the $N$ pixels.
  By using the bootstrap method, we also calculate a number
  $n_s$ of times the cross-correlation that each time chooses the same
  number $N$ of pixels from the original sample, but randomly selected
  (so that there are some pixels that are selected several times, while
  others are not selected at all). Both in bootstrap and jack-knife,
  we then calculate the r.m.s.  of these $n_s$ re-samplings, which provides our
  error. We use $n_s=10$, which implies that the relative
  error in the r.m.s. is $(2n_s)^{-1/2}\sim 20$\% for Gaussian
  errors. We note that for both
  techniques the $n_s$ determination have been performed on
  overlapping sub-samples, and they are thus not independent.

\item Field-to-field fluctuations: these are caused by intrinsic fluctuations in 
both the 
large-scale structure of galaxies and the microwave temperature field.
We propose three methods for estimating these fluctuations:

\begin{enumerate}

\item Different fields: We cross-correlate the $G$ field in the full
  area with a different field $\delta T_*$ of the same power
  spectrum as the original WMAP data, although uncorrelated with
  $G$. One simple way of
  applying this method is assigning to $\delta T_*$ the value of
  its own WMAP data but in other regions of the sky that are completely
  separated. For instance, we define $\delta T_*(l,b)=\delta T (l+\beta ,-b)$
  with different values of $\beta $ (we consider $n_s=10$ different
  values: $\beta =2i\pi /n_s$, $i$=1 to $n_s$), and calculate the
  r.m.s. for the $n_s$ realizations.
  In this case, we use a small enough number ($n_s$) of regions,
so the relative error in the r.m.s. is $(2n_s)^{-1/2}\sim $20\% for
Gaussian errors. The cross-correlations at scales 60-180$^\circ $ might 
produce some signal, but this would be small, given that the self-correlation 
of $\delta T$ is almost zero for $\theta >60^{\circ}$ \cite{copi}. 
The possible large-scale cross-correlations of the different fields 
infer that this estimation of the r.m.s. value is a conservative upper-limit value. 
  
\item Monte Carlo simulations of WMAP: We generate a
  number of Monte Carlo realizations of WMAP by using
  the software ``synfast'' to generate random mock maps of anisotropies
  corresponding to the theoretical power spectrum (Hinshaw et al. 2009)
  filtered for the V-band. We perform $n_S=100$ realizations, and then
  calculate the r.m.s. of their cross-correlation with the fixed 
  SDSS galaxy counts map. The relative error in the r.m.s. is $(2n_s)^{-1/2}=7$\%.

\item Integral of the self-correlations:
A calculation of the field-to-field variance in the cross-correlation 
of two non-correlated fields can be given by 
(see Appendix \S \ref{.bet})

\begin{equation}
\sigma ^2_{\omega _{TG}}(\theta)=
2\langle \omega _{TT}(\theta _{1,2}) \omega _{GG}
(\theta _{3,4})\rangle _{4,\theta }\ ,
\label{betancort}
\end{equation}
 where $\langle (...)\rangle _{4,\theta }$ stands for the average 
 extended over all groups of four pixels (1,2,3,4) in a region 
 in which the separation between pixels 1,3 and 2,4 is between $\theta -\Delta
 \theta /2$ and $\theta +\Delta \theta /2$,  $\theta _{1,2}$ is the 
 separation of pixels 1,2, $\theta _{3,4}$ is the 
 separation of pixels 3,4, and $\omega _{TT}$ and $\omega _{GG}$ are the
 self-correlations, respectively, for the fields $\delta _T$ and $\delta _G$. 
We note that with this method we assume that $\delta _T$ and $\delta _G$
are uncorrelated (as in Monte Carlo simulations); 
therefore, $\sigma $ refers to the limits 
of pure non-correlated fields within the corresponding probabilities (68\%). 
In addition, we note that we use the self-correlations that we
measure in our fields (see Fig. \ref{Fig:selfcorr}), i.e., we have only one realization. 
Cosmic variance would introduce some extra uncertainty.

\end{enumerate}

\end{enumerate}

\section{Results} 

\begin{figure}
\vspace{1cm}
{
\par\centering \resizebox*{6.5cm}{6.5cm}{\includegraphics{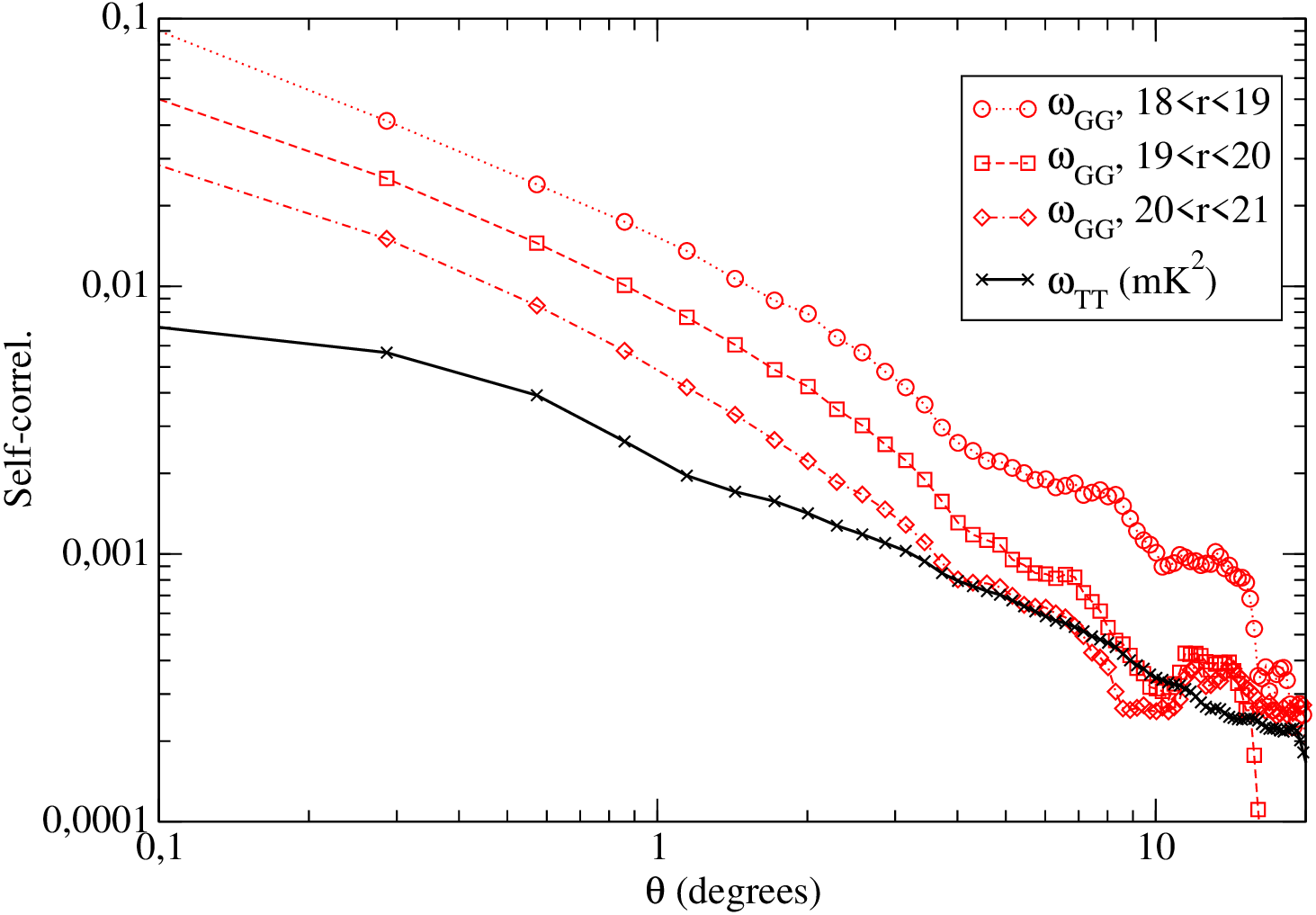}}
\par\centering
}
\caption{Log-log of the self-correlations of the fields $\delta _G$ 
and $\delta _T$.}
\label{Fig:selfcorr}
\end{figure}

\begin{figure*}[htb]
\vspace{1cm}
{\par\centering \resizebox*{5cm}{5cm}{\includegraphics{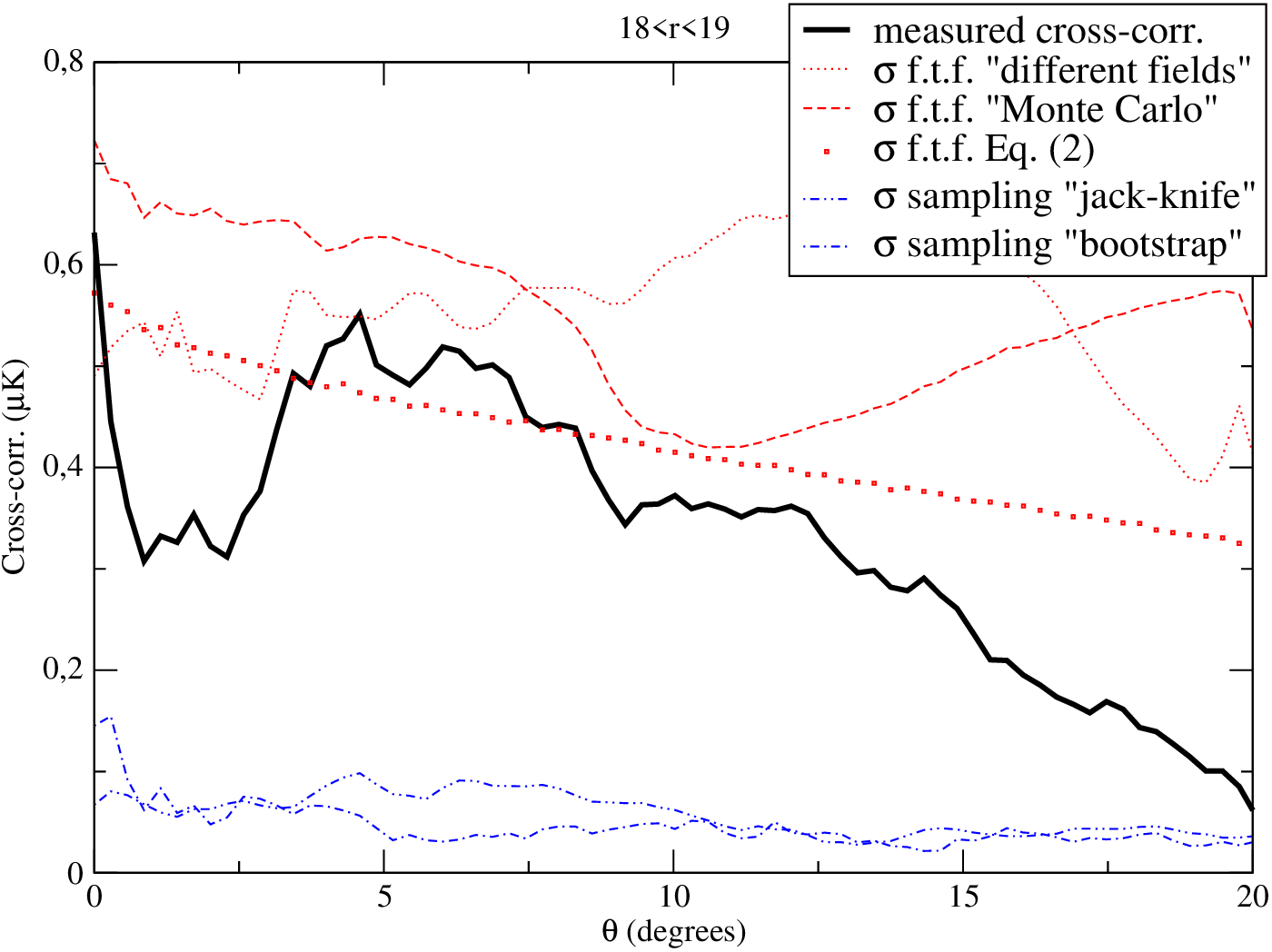}}
\hspace{.5cm}\resizebox*{5cm}{5cm}{\includegraphics{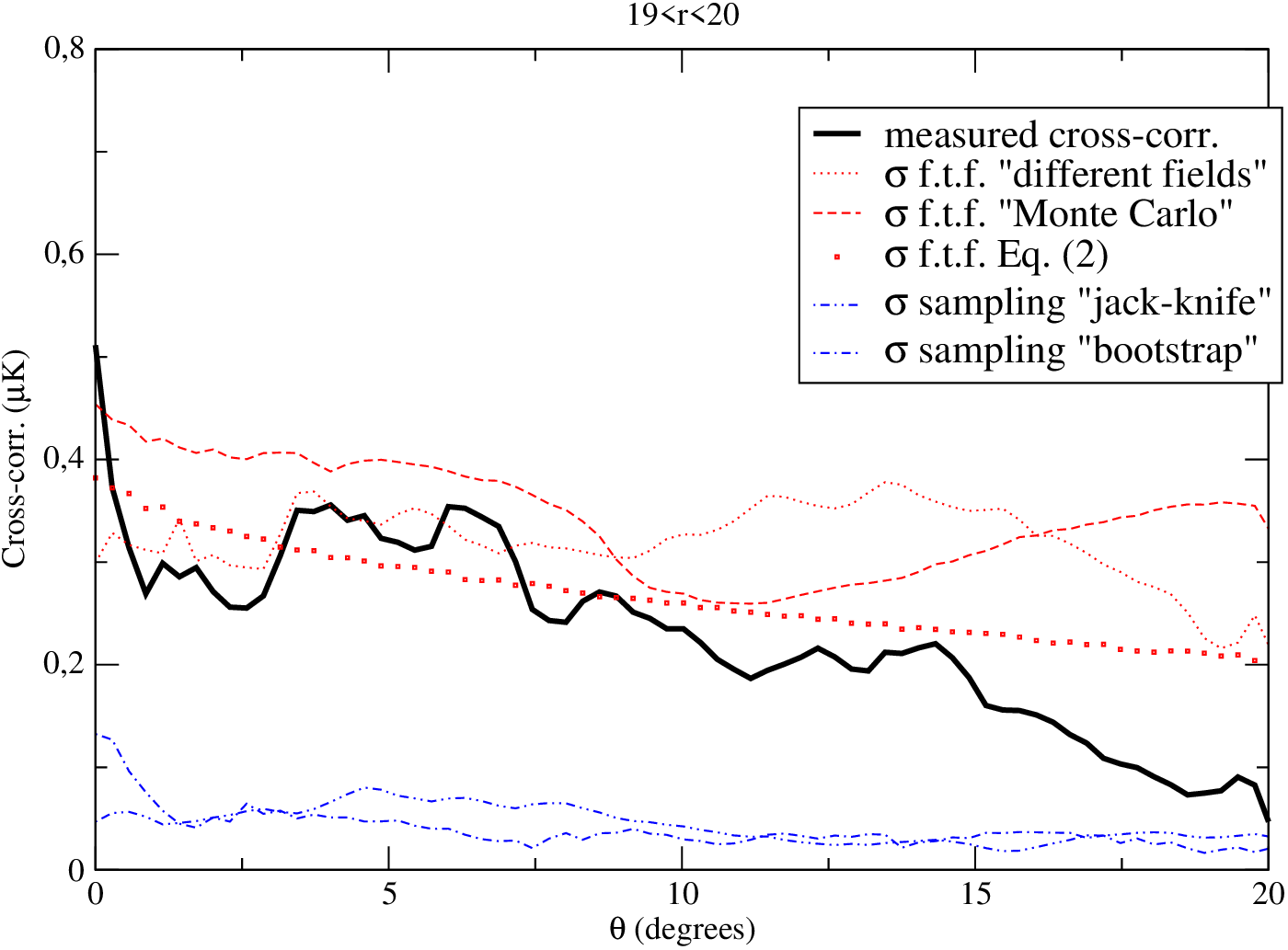}}
\vspace{.6cm}
\hspace{.5cm}\resizebox*{5cm}{5cm}{\includegraphics{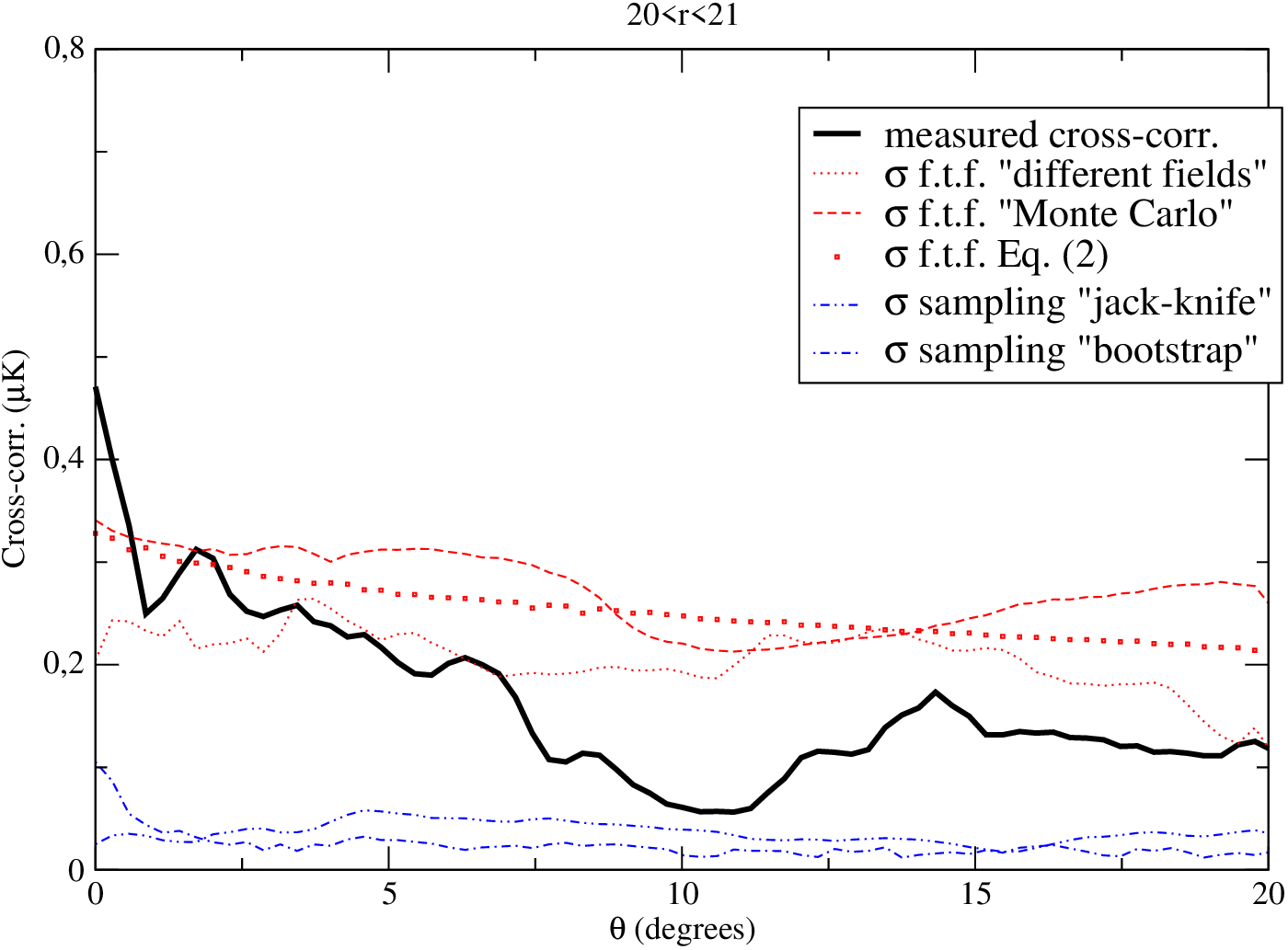}}\par}
\caption{
Cross-correlation function $\omega_{TG}$  WMAP-SDSS (black line)  for
galaxies with $b<30^{\circ}$ and
 in the magnitude range  $18 < r < 19$ (left panel),  $19 < r < 20$  (center
panel), and $20 < r < 21$ (right panel). The rms value
calculated by resampling errors and field-to-field fluctuations are
also plotted.}
\label{Fig:crosscorr}
\end{figure*}

In Fig. \ref{Fig:selfcorr}, we plot the self-correlations.
In Fig. \ref{Fig:crosscorr}, we show the determination of the
cross-correlation function for different ranges of magnitude, and the
errors computed by using re-sampling errors and field-to-field 
determinations. On the one hand, 
the errors computed by both the bootstrap and the jack-knife method are 
of the same order, and on the other hand the three  
``field-to-field'' methods yield similar results, 
which however are much larger than the re-sampling errors. The ``Different
fields'' method yields in general a slightly lower r.m.s. than the integral of
the self-correlations, possibly because of
small positive large-scale correlations, which slightly reduce the dispersion, as 
mentioned in discussing ``different fields'' in \S \ref{.methods}. 
The ``Monte Carlo'' method might yield slightly higher values of r.m.s. than
the integral of the self-correlations due to the larger amplitude of the low-multipoles in the theoretical 
power spectrum.

\begin{figure}[htb]
\vspace{1cm}
{
\par\centering \resizebox*{6.5cm}{6.5cm}{\includegraphics{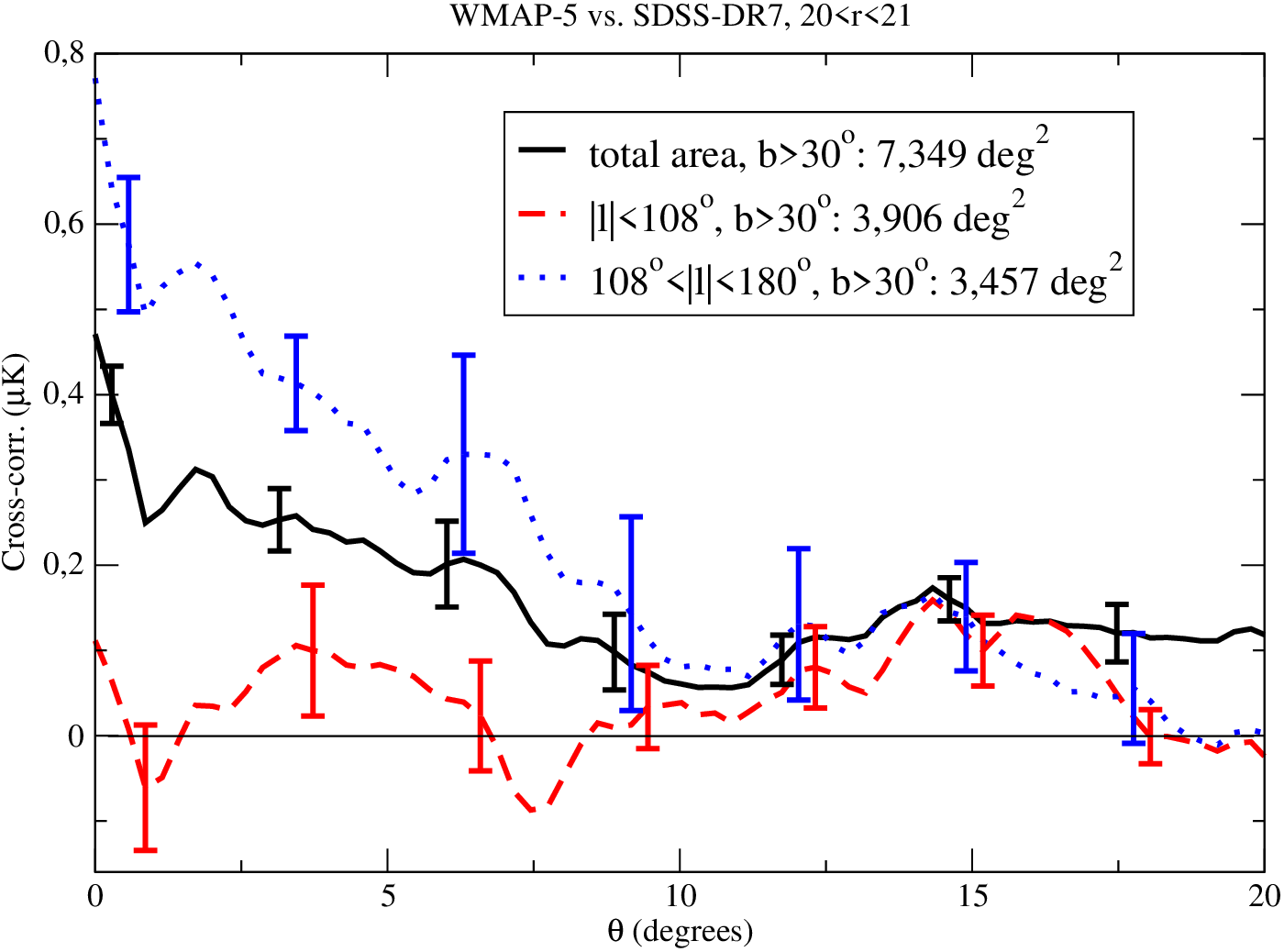}}
\par\centering
}
\caption{Cross-correlation ($\omega _{TG}$) WMAP-SDSS, $20<r<21$:
the  average of  the whole selected SDSS-DR7 area, and the average for 
  $|l|<108^\circ $ and $108^\circ<|l|<180^\circ $; error bars represent 
jack-knife re-sampling errors.}
\label{Fig:crosscorr2}
\end{figure}

The field-to-field fluctuations obtained by using independent determinations
of the cross-correlation function are similar to the amplitude of the detected signal or
even larger. Figure \ref{Fig:crosscorr2} illustrates 
this point by showing
that there are no positive average  cross-correlations in a 
sky region of area more than half of the full angular coverage.

From all these analyses, we cannot exclude the value of $|\omega
_{TG}|$ being compatible with zero for any $\theta $ within
field-to-field fluctuations.
Thus we conclude that there is no significant
cross-correlation detection. This situation is similar to
that found for the SDSS 3D self-correlation by Sylos-Labini et al. (2009), 
who also demonstrated that the field-to-field fluctuations are of the 
order of the signal in the previously announced discovery of 
baryon acoustic oscillations and large-scale anticorrelations.

\section{Comparison with previous works}

Other authors who calculated the
cross-correlation between WMAP and SDSS galaxy counts measured 
a significant signal.  For instance, Cabr\'e et al. (2006) 
measured a value of $\omega _{TG}(3^\circ )\sim 0.5$ $\mu $K and a significant
positive $\omega _{TG}$ for all angles lower than 20 degrees for the
subsample $20<r<21$ over 5500 square degrees of SDSS-DR4. 
Giannantonio et al. (2008) obtained a value of $\omega _{TG}(3^\circ )\sim
0.3$ $\mu $K for the subsample $18<r<21$ of SDSS-DR6, excluding
the southern Galactic hemisphere and high Galactic extinction regions. In addition,
they found a significant positive signal out to $\theta =8$
degrees. Their values are more or less compatible with
our estimate of the
cross-correlation function, within the re-sampling error bars and taking
into account that their subsamples are slightly different. However  
we do not measure significant cross-correlations, whereas these author do
a result we cannot explain.

Cabr\'e et al. (2006) and
Giannantonio et al. (2008) performed Monte Carlo simulations using mock maps, and obtained similar values or ones only slightly larger than a jack-knife. We do not know whether these disagreement are caused by
mistakes in their calculations or whether their claim is 
that re-sampling errors represent the full errors. Other authors used only
jack-knife technique errors (e.g., Sawangwit et al. 2009).  A similar problem may affect
the results of Raccanelli et al. (2008), who measured the
cross-correlation between NVSS radio sources and WMAP anisotropies.
Raccanelli et al. (2008) calculate the error in simulating 1,000 mock NVSS
maps by randomly distributing the unmasked pixels of the true NVSS
maps.  We are concerned that this process might destroy part of the
self-correlation of each map, and that the errors might not represent the full 
field-to-field fluctuations.  There has been considerable discussion
of these errors \cite{Cabre07,Giannantonio08}. 
However, against their claims one can infer from the
analyses of this paper that: i) jack-knife or bootstrap methods do not
calculate the whole error; ii) the level field-to-field fluctuations
is as large as the measured average signal.
In addition, our conclusion is that the signal is 
largely dependent on the specific sub-region chosen. We find
that in the large area of $|l|<108^\circ $, $b>30^\circ $ 
(3,906 square degrees available with SDSS-DR7, more than half of the sample) 
we do not measure any signal, so the average signal of the entire sample must be 
caused by a fluctuation.

One remarkable aspect of the analysis of WMAP/SDSS-DR4 is that 
Cabr\'e et al. (2006) obtain a 3.6$\sigma $ detection for $20<r<21$, while
Giannantonio et al. (2008) with a wider areal coverage (SDSS-DR6) 
and broader range of magnitudes, $18<r<21$, achieve only a 2.3$\sigma $ 
detection. This decrease in the significance is unexpected if the 
signal were real. 
We also note that some authors calculated the combined signal to noise 
ratio of different cross-correlations in different samples, obtaining values 
over 4$\sigma $, by summing them quadratically \cite{Cabre06}. 
This is incorrect because they do not take into account the correlation 
between the samples, thus neglect an important part of the estimated error.

A higher significance in the cross-correlation of
WMAP/SDSS is claimed to be obtained
\cite{Granett08} when only super-clusters/super-voids are correlated
with WMAP instead of the entire SDSS survey: a value of 4.4$\sigma $. 
Apart from our questions raised above, we are also concerned about
possible a posteriori fitted parameters used to obtain this
correlation. For instance, Granett et al. (2008) separate regions of
the sky centered on super-clusters with radii of 4 degrees; we ask why 4
degrees? These authors illustrate that the significance is only
3.5$\sigma $ for radius 3 deg or 3.8$\sigma $ for 5 deg. The significance
also changes with the number of superclusters/supervoids selected, being
4.4$\sigma $ with $N=50$ but only 2.8$\sigma $ with $N=70$.
A signal to noise ratio of 2-3$\sigma $ is provided by other authors without any selection of super-clusters, so they appear to have considered both a 
radius and number of
superclusters/supervoids that achieves the maximum increase in the
signal to noise ratio (from 2-3 up to to 4.4).

We note that Bielby et al. (2009) measured the correlation of WMAP
anisotropies with emission-line galaxies selected photometrically from
SDSS and inferred a non-significant correlation, with large 
field-to-field errors comparable to those we obtain
(we are cautious in interpreting their result, however, because
the cross-correlation in the different subfields are not independent and
this affects the way in which they have been using to determine the r.m.s.). 
They claim that their result implies that
possibly emission line galaxies are more strongly clustered and less
correlated with microwave anisotropies, something that is not
entirely clear to us. In our opinion, the results of Bielby et al. (2009)
of non-significant cross-correlation may be correct and
there is unlikely to be a difference in the interaction of galaxies with
the background CMBR that is caused entirely by them having emission lines.
Sawangwit et al. (2009) failed to measure a significant 
cross-correlation between the luminous red galaxies of SDSS-DR5 and WMAP. 
On the other hand, they found some positive correlation of WMAP with 
2SLAQ survey, and negative correlation of WMAP with the AAOmega survey. 
Analyses by Hern\'andez-Monteagudo (2008) demonstrated that
the cross-correlation of WMAP/SDSS-DR4 should have at least within a signal/noise ratio 
of 0.7-1.7, much lower than the significance obtained by the authors 
cited above. These results might be interpreted as independent 
confirmations of our results here.

Apart from those analyzing data from SDSS, previous studies reach 
the general conclusion that the ISW effect was not detected significantly
in: (1) cross-correlations with X-ray XRB, Boughn \& Crittenden (2003)
claiming an absence of the ISW using X-ray data; (2) near-infrared 2MASS,
Francis \& Peacock (2009) not finding any corresponding ISW signal; or (3)
radio sources NVSS, Hern\'andez-Monteagudo (2009) casting doubt on the correlation
between WMAP and NVSS radio sources, since the cumulative
signal to noise ratio of the cross-correlation with multipoles $l<60$ is lower
than 1, and the ISW itself, since the signal to noise ratio should be around 5 theoretically.

\section{Conclusions}

We concluded that there is no significant cross-correlation
between the CMBR temperature anisotropies of WMAP and the galaxy counts of 
the SDSS, and any claims to have detected the ISW 
effect on the basis of significant cross-correlation are unjustified. 
Field-to-field fluctuations
dominate the detected signals. Any detection of signal is very
dependent on the selected region of the sky. 
Other authors erroneously claimed to have detect significant correlations 
because they had used a particular sky region with a fluctuation that 
is not representative of the average sky or because 
they had underestimated the statistical errors by using 
non-independent re-samplings. If our conclusion is correct, the value of 
$\Omega _\Lambda\approx 0.8$ obtained by those
authors based on the assumption of observing the ISW effect 
would have been one induced by noise. Its value would be 
coincident with the expected value for 
$\Lambda $CDM by chance, and in the spirit of accepting a 
scientific result when it indeed produces numbers expected a priori.

\acknowledgements   
We thank an anonymous referee for useful comments and suggestions.
Thanks are given
to J. A. Rubi\~no-Mart\'\i n and R. G\'enova-Santos from IAC (Tenerife) 
for helpful comments and help in the use of HEALPIX and SYNFAST software.
Thanks are given to Claire Halliday (language editor of A\&A) for proof-reading of the text. Funding for the SDSS has been provided by the 
Alfred P. Sloan Foundation, the Participating
Institutions, the National Science Foundation, the U.S. Department of
Energy, the NASA, the Japanese Monbukagakusho, the Max Planck Society,
and the Higher Education Funding Council for England. The SDSS Web
Site is: http.//www.sdss.org/. WMAP is the result of a partnership
between Princeton University and NASA's Goddard Space Flight Center.
MLC was supported by the {\it Ram\'on y Cajal} Programme of the 
Spanish Science Ministry.

{}

\appendix

\section{Field-to-field errors in the cross-correlation 
as an integral of the self-correlations for two uncorrelated 
fields} \footnote{By J. Betancort-Rijo.}
\label{.bet}

We consider two continuous random scalar fields, $F_A$ and $F_B$, in
a space with $d$ dimensions and any topology. Without loss of generality,
we shall assume that the mean values (over realization) of both fields 
is zero and they are not correlated:
$\langle F_A\rangle=\langle F_B\rangle =\langle F_AF_B\rangle=0$.
On the other hand, the values of each field at two different points
$\vec{x}_1$, $\vec{x}_2$ are not independent random variables:  
$\omega _{A/B}(\vec{x}_1,\vec{x}_2)=
\langle F_{A/B}(\vec{x}_1)F_{A/B}(\vec{x}_2)\rangle$, where the average is over
realizations. In practice, in most interesting cases the fields are
statistically homogeneous and ergodic, so that $\omega $ depends only
on $\vec{x}_1-\vec{x}_2$, and the correlation may be defined as
spatial averages, which is the useful definition since in most cases
only one realization is available. If the fields are also statistically
isotropic, $\omega $ depends only on $r\equiv |\vec{x}_1-\vec{x}_2|$.
For the following derivation, we shall assume homogeneity and
isotropy; the full expression might easily be recovered if needed.

We first derive the ``field-to-field error'' (i.e., the true error) 
for the zero lag estimator:
\begin{equation}
E[\omega _{AB}(r=0)]=\frac{1}{N}\sum _{i=1}^NF_A(i)F_B(i)
\label{bet1}
,\end{equation}
where $F_{A/B}(i)\equiv F_{A/B}(\vec{x}_i)$. We have replaced the
$d$-dimensional volume integral 
over the sample with a sum over $N$ equal volume cell indexed by $i$
and centered on $\vec{x}_i$. We would have to multiply the contribution of
the field for each pixel by a weight equal to the volume of the pixel in the
case of non-equal volume cells.

For the variance in Eq. (\ref{bet1}), we have:
\begin{equation}
\sigma ^2_{\omega _{AB}}(r=0)=\langle E^2\rangle-\langle E\rangle ^2=
\langle E^2\rangle
,\end{equation}
since, by construction, the mean value of $E$ over realizations, $\omega _{AB}$,
is assumed to be zero. Developing the square of expression (\ref{bet1}), and
taking its average, we have:
\begin{equation}
\langle E^2\rangle=\frac{2}{N^2}\left\langle \sum _{i,j}^N F_A(i)F_B(i)
F_A(j)F_B(j)\right\rangle
.\end{equation}
In principle, the factor of 2 should not be 
there in the case $i=j$, but this will
be negligible in the limit of arbitrarily small cells.

Now, since the fields $F_A$ and $F_B$ are uncorrelated
\begin{equation}
\langle F_A(i)F_B(i)F_A(j)F_B(j)\rangle=\langle F_A(i)F_A(j)\rangle
\langle F_B(i)F_B(j)\rangle =\omega _A(r_{ij})\omega _B(r_{ij})
,\end{equation}
where $r_{ij}\equiv |\vec{x}_i-\vec{x}_j|$. Thus, we have
\begin{equation}
\sigma ^2_{\omega _{AB}}(r=0)=2\langle \omega _A(r)\omega _B(r)\rangle
\equiv \frac{2}{V_s^2}\int \int _{\rm sample}\omega _A(r_{12})
\omega _B(r_{12}) d^d\vec{r}_1d^d\vec{r}_2
\label{bet5}
,\end{equation}
where $V_s$ represents the volume of the sample.

The correlation estimator for any non-zero lag is
\begin{equation}
E[\omega _{AB}(r_0)]=\frac{1}{N^2}\sum _{i,j/ r_{ij}=r}^N F_A(i)F_B(j)
\label{bet6}
,\end{equation}
where $r_0-\Delta r/2<r<r_0+\Delta r/2$.
Following Eq. (\ref{bet6}) using the same procedure as for Eq. (\ref{bet1}),
one obtains
\[
\sigma ^2_{\omega _{AB}}(r_0)=2\langle \omega _A(r_{12})\omega _B(r_{34})
\rangle _{4,r_0}\equiv 
\]\begin{equation}
\frac{2\int \int _{{\rm sample}/ r_{13}=r,r_{24}=r}
\omega _A(r_{12})\omega _B(r_{34}) 
d^d\vec{r}_1d^d\vec{r}_2d^d\vec{r}_3d^d\vec{r}_4}
{\int \int _{{\rm sample}/ r_{13}=r,r_{24}=r} 
d^d\vec{r}_1d^d\vec{r}_2d^d\vec{r}_3d^d\vec{r}_4}
\label{bet7}
.\end{equation}

Equation (\ref{bet7}), and its particular case, Eq. (\ref{bet5}),
infer the variance over realizations of the estimator of the correlation
between two uncorrelated fields $A$, $B$ when any new global realization of
both fields is carried out. In the case when we fix the realization of
one of the fields while changing the other, Eq. (\ref{bet7}) is also
valid but the self-correlation of the fixed realization must be calculated
by averaging over pixels in this fixed realization, rather than over realizations.
Now, since the estimated self-correlation may fluctuate above and below the
universal (mean of all realizations) value, it is clear that the variance in
the estimator of the cross-correlation of $A$ and $B$ when one of them
is kept fixed may be slightly above or below the one corresponding to
the case when both fields fluctuate.

\end{document}